\documentclass[draftclsnofoot,12pt,onecolumn,a4paper]{IEEEtran}

\usepackage {graphicx}
\usepackage {amsmath}
\usepackage {ulem}
\usepackage {latexsym}
\usepackage {multirow}
\usepackage{array}
\usepackage[dvips]{epsfig}
\usepackage{subfigure}
\usepackage {enumerate}

\begin{document}

\title{Interference Mitigation Using Uplink Power Control for Two-Tier Femtocell Networks}
\author{Han-Shin Jo,~\IEEEmembership{Student Member,~IEEE,}
        Cheol Mun,~\IEEEmembership{Member,~IEEE,}
        June Moon,~\IEEEmembership{Member,~IEEE,}
        and Jong-Gwan Yook~\IEEEmembership{Member,~IEEE}
\thanks{Han-Shin Jo and Jong-Gwan Yook are with Department of Electrical \& Electronic Engeering, Yonsei University,
        Seoul, Korea 120-149. (e-mail: \{gminor, jgyook\}@yonsei.ac.kr).}
\thanks{Cheol Mun is with the Dept. of Electronic Communication
Eng., Chungju National University, Chungju, Korea.
(e-mail:chmun@cjnu.ac.kr).}
\thanks{June Moon is with Telecommunication R\&D Center, Samsung Electronics, Suwon, Gyeonggi, Korea
442-742. (e-mail:june.moon@samsung.com).}
\thanks{This work was supported by Samsung Electronics.}}

\markboth{Submitted to IEEE Transactions on Wireless
Communications}{Shell \MakeLowercase{\textit{et al.}}: Bare Demo of
IEEEtran.cls for Journals} \maketitle

\begin{abstract}
This paper proposes two interference mitigation strategies that
adjust the maximum transmit power of femtocell users to suppress the
cross-tier interference at a macrocell base station (BS). The
open-loop and the closed-loop control suppress the cross-tier
interference less than a fixed threshold and an adaptive threshold
based on the noise and interference (NI) level at the macrocell BS,
respectively. Simulation results show that both schemes effectively
compensate the uplink throughput degradation of the macrocell BS due to the cross-tier interference and
that the closed-loop control provides better femtocell throughput
than the open-loop control at a minimal cost of macrocell
throughput.
\end{abstract}


\IEEEpeerreviewmaketitle

\section{Introduction}
Femtocells have emerged as a solution to increase both the capacity
and coverage while reducing both capital expenditures and operating
expenses. Femtocells, which consist of miniature personal base
stations and stationary or low-mobility end users deployed in an
indoor environment, are located within an existing cellular network;
a two-tier network is deployed. Since femtocells operate in the
licensed spectrum owned by wireless operators and share this
spectrum with macrocell networks, limiting the cross-tier
interference from femtocell users at a macrocell base station (BS)
is an indispensable condition for deploying femtocells in the
uplink. In addition, since a network operator can not control the
femtocell locations, femtocells need to sense the radio environments
around them and carry out self-configuration and self-optimization
of radio resources from the moment they are set up in by the
consumer \cite{SOC1}.

Previous studies have developed cross-tier interference control
methods by using transmit power control
\cite{OverlayHotspot},\cite{OverlayMicro} and time hopping coupled
with antenna sectoring \cite{Fetocell} for a two-tier CDMA network.
A signal-to-interference ratio (SIR) based power control scheme is shown to promote uplink performance, but does not
consider the wall penetration loss \cite{OverlayHotspot}. On the other hand, under the assumption that macrocell
BS knows the positions of the overlaid users, a power control scheme has been proposed
to minimize the total transmitted power while satisfying the SIR
requirement and considering the wall penetration loss \cite{OverlayMicro}.
The schemes in \cite{OverlayHotspot} and \cite{OverlayMicro} have
different effects on the uplink performances according to the
underlying femtocell locations; they require site-specific
engineering in which the system parameters are manually tuned and
optimized by technicians. Therefore, these methods are not
applicable to femtocells that should carry out the automatic self-configuration of radio
resources.

This paper proposes two interference mitigation strategies in which
femtocell users adjust the maximum transmit power using an open-loop
and a closed-loop technique. With the open-loop control, a femtocell
user adjusts the maximum transmit power to suppress the cross-tier
interference due to the femtocell user, which results in the
cross-tier interference less than a fixed interference threshold.
The closed-loop control adjusts the maximum transmit power of the femtocell user to
satisfy an adaptive interference threshold based on the level of
noise and uplink interference (NI) at the macrocell BS. Both schemes
effectively compensate the uplink throughput degradation of the
existing macrocell BS due to the cross-tier interference.
Furthermore, the use of the closed-loop control provides femtocell
throughput that is superior to that using open-loop control at a
very low macrocell throughput cost.

The remainder of the paper is organized as follows. The two-tier
cellular system model and basic power control method is described in
Section~\ref{sec:model}. Section~\ref{sec:TPC} is devoted to
describing the proposed open-loop and closed-loop control schemes
for the maximum transmit power of femtocell users. In Section~\ref{sec:results}, the
simulation results are presented and discussed. Finally, the
conclusions are drawn in Section~\ref{sec:conclusion}.

\section{System Model}
\label{sec:model} A two-tier CDMA cellular network is considered
here. A macrocell has a layout of two rings, 19 cells and single
sector per cell. In-building femtocells are uniformly distributed in
a macrocell. Each femtocell BS is located within the center of the
building and provides indoor wireless coverage to femtocell users.
Macrocell and femtocell users are uniformly distributed.

The radio links between the BS and the users are divided into outdoor, indoor,
and outdoor-to-indoor links according to the radio environment. The
propagation loss $L$ caused by slow fading at the uplink is modeled
based on an ITU and COST231 path loss model
\cite{ITUPL}\cite{Cost231}:
\begin{itemize}
\item Outdoor link (macrocell user $\rightarrow$ macrocell BS)
\begin{equation}
L=10^{4.9}\left(\frac{d}{1000}\right)^{4}f^{3}10^{S/10},\label{eq:PLout}
\end{equation}
\item Outdoor-to-indoor link (macrocell user $\rightarrow$ femtocell BS, femtocell user $\rightarrow$ macrocell BS)
\begin{equation}
L=10^{4.9}\left(\frac{d}{1000}\right)^{4}f^{3}10^{S/10}
10^{(L_i+L_e)/10}.\label{eq:PLoutin}
\end{equation}
\item Indoor link (femtocell user $\rightarrow$ femtocell BS)
\begin{equation}
L=10^{3}d^{3.7}10^{S/10}10^{L_i/10},\label{eq:PLin}
\end{equation}
\end{itemize}
Here, $d$ and $f$ are the amount of transmitter-receiver separation
in meter and the frequency in MHz, respectively. $S$ represents
log-normal shadowing with a standard deviation of 8 dB and a
shadowing correlation of 0.5 in (\ref{eq:PLout}) and
(\ref{eq:PLoutin}), and that with a standard deviation of 10 dB and
a shadowing correlation of 0.7 dB in (\ref{eq:PLin}). $L_e$ denotes
the external wall loss modeled as a Gaussian distribution with a
mean of 7 dB and a standard deviation of 6 dB. $L_i$ is internal
wall loss given by $L_i=4I$, where $I$ is Bernoulli random variable
with success parameter $p=0.5$.

%

In the conventional uplink open-loop power control, the uplink
path loss (including shadowing) between a user and a macrocell BS can be estimated by
both the received downlink signal power measured at the user and the broadcast
information on the effective isotropic radiated power (EIRP) of the
macrocell BS. With further knowledge of the required
signal-to-interference-and-noise ratio (SINR) of the current
transmission and with broadcast information on the average power
level of the noise and interference at the BS, the required uplink
transmit power $P_r$ is derived as follows:
\begin{equation}
P_{r}=L\cdot NI\cdot\gamma_0,\label{eq:ULTxPw1}
\end{equation}
Here, $L$, $\gamma_0$, and $NI$ are the uplink propagation loss, the
required SINR for the current modulation and coding rate, and the
estimated average power level of the noise and interference at the
BS without considering the BS antenna
gain, respectively. 
Finally, the transmit power of the user $P_t=\left(P_{r},
\overline{P}_{max}\right)$ is determined not to be in excess of the
maximum transmit power $\overline{P}_{max}$, where the maximum
transmit power $\overline{P}_{max}$ can not be adaptively adjusted
because it is the fixed value satisfying the spectrum emission mask
and the error vector magnitude requirement.

In this paper, the upper limit of transmit power $P_{max}$
($\leq\overline{P}_{max}$) is introduced to control the cross-tier
interference caused by femtocell users, and the transmit power of
the user is determined not to be in excess of $P_{max}$ as follows:
\begin{equation}
P_t=\min\left(P_{r}, P_{max}\right).\label{eq:ULTxPw2}
\end{equation}
The additional cross-tier interference at a macrocell BS due to
femtocell users leads to a higher $P_r$ as shown in
(\ref{eq:ULTxPw1}), which leads to a NI rise at the macrocell BS,
which results in uplink throughput degradation in the macrocell.
Therefore, adjusting $P_{max}$ according to the cross-tier
interference level can limit the NI rise and thus compensate for the
uplink throughput degradation in the macrocell. In the following
chapters, the maximum transmit power is used for naming $P_{max}$.

\section{Uplink Power Control in a Femtocell}

In this section, an uplink power control scheme for femtocell users
is proposed that adjusts the maximum transmit power $P_{max}$ as a
function of the cross-tier interference level in an open-loop and
closed-loop technique.

\label{sec:TPC}
\subsection{Open-Loop Control for $P_{max}$}

In the proposed open-loop control for $P_{max}$, a femtocell user estimates its additional cross-tier interference to
the macrocell BS and adjusts the maximum transmit power $P_{max}$
such that the additional cross-tier interference power is less than
the predetermined maximum allowable interference level.

A femtocell user estimates the received signal powers from $K$
macrocell BSs configured in its neighbor BS
list\footnote{The neighbor BS list of each femtocell, which is
auto-configured by its serving femtocell BS, contains several macrocell and femtocell BSs
that cause the strong interference to the femtocell. Therefore, even
for a realistic scenario where there are many femtocells deployed, a
femtocell user completely find the neighbor macrocell BSs that are affected most by
the cross-tier interference without monitoring all neighbor BSs.} $\{R_k\}_{k=1,\cdots,K}$, and evaluates the propagation losses
$\{L_{k}\}_{k=1,\cdots,K}$ from the macrocell BSs as follows:
\begin{equation}
L_{k}=\frac{E_{k}}{R_{k}},\label{eq:LinkLoss}
\end{equation}
where the EIRP of the $k$th macrocell BS, $E_{k}$ is obtained by the broadcast information from the BSs. The propagation losses $L_{k}$ can consider the path loss (including shadowing) in a frequency division duplexing (FDD) system by using the received signal powers $\{R_k\}_{k=1,\cdots,K}$ averaged over time. In a time division duplexing (TDD) system, fast fading can be also considered in $L_{k}$ by using the instantaneous received signal powers $\{R_k\}_{k=1,\cdots,K}$ due to the channel reciprocity between uplink and downlink.

Based on the symmetry between the uplink and downlink propagation losses, the $k^\ast$th
macrocell BS with the minimum propagation loss $L_{\min}=\min(L_1,L_2,\cdots,L_{K})$ is affected most by the
cross-tier inference of the femtocell user. Thus, the maximum cross-tier inference of the femtocell user to the $k^\ast$th macrocell BS,
$I_{k^\ast}$ should satisfy the interference power constraint as follows:
\begin{equation}
I_{k^\ast}=\frac{P_{t}}{L_{min}}\leq I_{th,k^\ast},~\textrm{and}~
I_{th,k^\ast}=\frac{\alpha N_0 W F}{J_{k^\ast}}.\label{eq:I}
\end{equation}
Here, $P_t$ is the transmit power of the femtocell user.
$I_{th,k^\ast}$ is the maximum allowable cross-tier interference
power at the $k^\ast$th macrocell BS.
$N_0$, $W$, and $F$ denote the thermal noise spectral density, the
bandwidth, and the noise figure of the macrocell BS, respectively,
and the interference-to-noise ratio $\alpha$ is a system-specific
parameter known a priori to femtocell users. ${J_{k^\ast}}$ denotes
the number of active femtocell users that decide the $k^\ast$th
macrocell BS is affected most by their own interferences. Using the
value of $k^\ast$ reported from each femtocell user, each macrocell
BS determines ${J_{k^\ast}}$ and broadcasts it to the femtocell
users via the femtocell BSs.

The maximum transmit power of the femtocell user determined by an
open-loop method, $P_{max,OL}$, is given by
\begin{equation}
P_{max,OL}=I_{th,k^\ast}\cdot L_{min}.\label{eq:OpenPmax_k}
\end{equation}
The transmit power of the femtocell user $P_t$ is determined by
(\ref{eq:ULTxPw2}), in which $P_{max}$ is replaced by $P_{max,OL}$.
This guarantees that the maximum cross-tier inference of the
femtocell user does not exceed a the maximum allowable cross-tier
interference power of each macrocell BS. Since the actual NI level
at each macrocell BS is not considered in the open-loop method,
however, even when the NI level at each macrocell BS is sufficiently
low, the maximum transmit power of femtocell users must be more
limited than necessary. This degrades the uplink throughput in the
femtocell.

\subsection{Closed-Loop Control for $P_{max}$}
In the proposed closed-loop control for $P_{max}$, a femtocell user
adjusts the maximum transmit power $P_{max}$ as a function of not
only the estimated additional cross-tier interference due to the
femtocell user but also the NI level at the macrocell BS. Before the
closed-loop control procedure begins, a femtocell user obtains
$\{R_k\}_{k=1,\cdots,K}$ and $\{E_k\}_{k=1,\cdots,K}$ and evaluates
$\{L_k\}_{k=1,\cdots,K}$ via (\ref{eq:LinkLoss}) using the same
method used in the open-loop control. Additionally, the $K$
macrocell BSs continuously monitor the uplink interference power
from both macrocell users and femtocell users and broadcast the NI
levels to the femtocell users via the femtocell BS that are wired to
the macrocell BSs.

\begin{figure}
\begin{center}
   \includegraphics[width=5.5in]{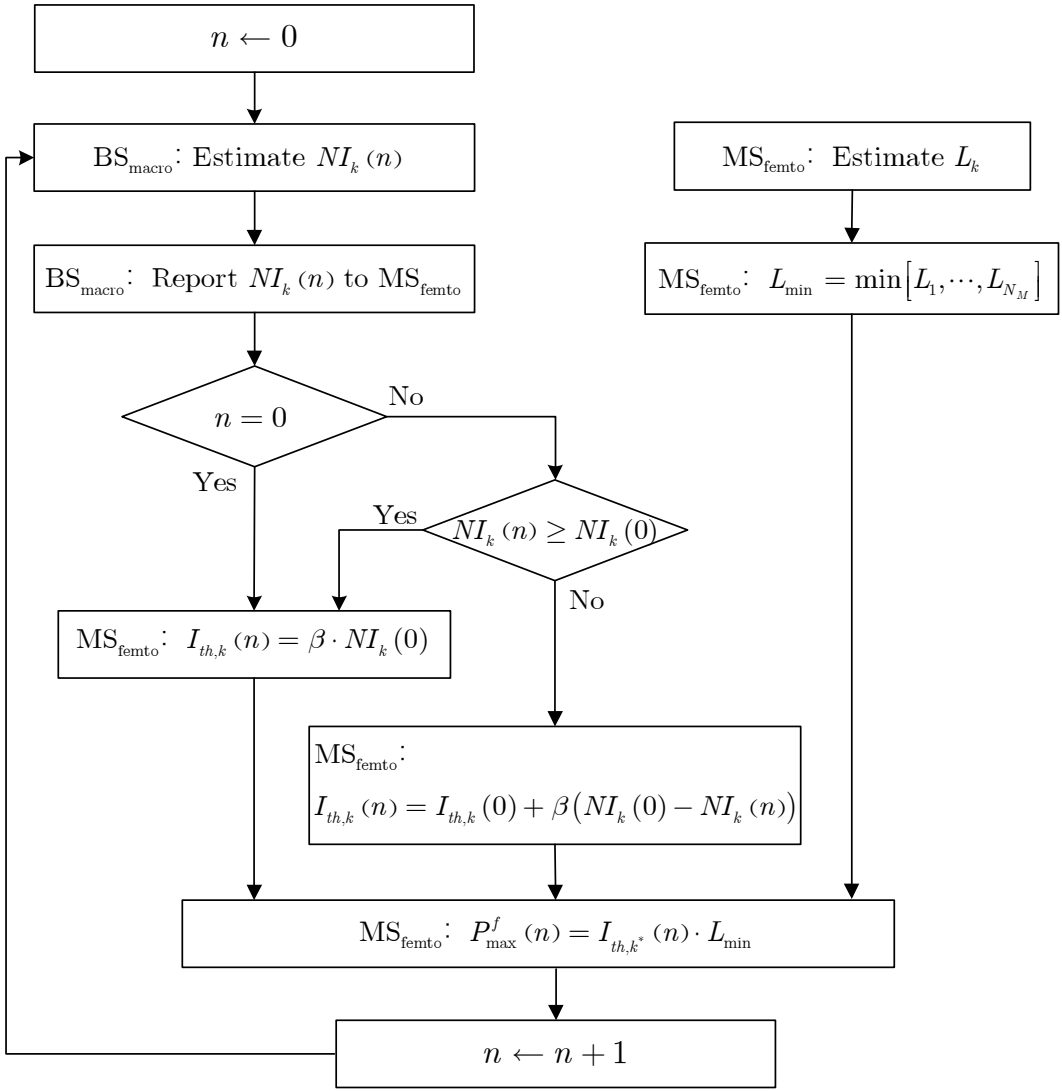}
    \caption{Flow diagram for the closed-loop power control scheme}
    \label{fig:Algorithm}
\end{center}
\end{figure}

Fig. \ref{fig:Algorithm} shows the detailed closed-loop procedure
used to update the maximum transmit power. For the initial step
$n=0$, before the femtocell user becomes active, the user obtains
the initial NI level of the $k$th macrocell BS, $NI_{k}(0)$ including
the uplink interference power from active users of macrocells and femtocells.

For $n>0$, after the femtocell user becomes active, $NI_{k}(n)$
includes the additional cross-tier interference due to the femtocell
user. This is given by
\begin{eqnarray}
NI_{k}(n) = NI_{k}(0) + \frac{T^f}{L_k^f},
\end{eqnarray}
where $T^{f}$ is the transmit power of the active femtocell user and
$L^f_k$ is the indoor-to-outdoor propagation loss from the femtocell
user to the $k$th macrocell BS.

The maximum allowable interference power of the $k$th macrocell BS,
$I_{th,k}(n)$ is determined as a function of $NI_k(n)$ and
$NI_k(0)$, as follows:
\begin{equation}
I_{th,k}(n) = \left\{ \begin{array}{ll}
\beta \cdot NI_{k}(0) & \textrm{$n=0$ or if $NI_k(n) \geq NI_k(0)$ for $n>0$}\\
I_{th,k}(0) + \beta \left( NI_k(0) - NI_k(n)\right) & \textrm{if
$NI_k(n) < NI_k(0)$ for $n>0$}
\end{array} \right.,
\end{equation}
where $\beta$ is a system-specific parameter known a priori to
femtocell users. If the NI level of the $k$th macrocell BS increases
due to the additional cross-tier inference of a new active femtocell
user, i.e., $NI_{k}(n) \geq NI_{k}(0)$, the maximum allowable
interference power, $I_{th,k}(n)$, is determined on the basis of the
NI level if there are no active femtocell users, $NI_{k}(0)$. This
decreases the transmit power of the femtocell user by lowering the
maximum transmit power. On the other hand, if the inference of
active macrocell users decreases, the NI level of the $k$th
macrocell BS decreases regardless of the additional cross-tier
inference of a new active femtocell user, i.e., $NI_{k}(n) <
NI_{k}(0)$. In this case, $I_{th,k}(n)$ increases to be
$\beta(NI_{k}(0)-NI_{k}(n))$ greater than $I_{th,k}(0)$, which
increases the maximum transmit power and thus improves the uplink
throughput in the femtocell.

The maximum transmit power of the femtocell user by the closed-loop
method, $P_{max,CL}$ is given by
\begin{equation}
P_{max,CL}(n)=I_{th,k^\ast}(n)\cdot L_{min},\label{eq:ClosePmax_k}
\end{equation}
where $I_{th,k^\ast}(n)$ is the maximum allowable interference power
of the $k^*$th macrocell BS with the minimum propagation loss
$L_{min}=\min(L_1,L_2,\cdots,L_{K})$. The transmit power of the
femtocell user $P_t$, which is determined using $P_{max,CL}$ from
(\ref{eq:ULTxPw2}), enhances the uplink throughput in the femtocell
while bringing the total NI level of the macrocell BS under control.

\section{Performance Evaluation}
\label{sec:results} System parameters based on CDMA2000 1xEV-DO
Revision A (DO Rev. A) \cite{EVDO} is used for the performance
evaluation. For a system occupying a bandwidth of 1.25 MHz with a
center frequency of 2.5 GHz, an omnidirectional antenna with a gain
of 0 dBi are used for both user and BS. At each drop, 10 macrocell
users are uniformly distributed in each sector with a $R$ of
$800/\sqrt{3}$ meters. They are connected to the macrocell BS with
the lowest path loss among all possible links to the macrocell BSs.
Four femtocell users are uniformly distributed in a building with a
width and length of 50 meters, and a single femtocell BS is located
in the center of the building. Each physical layer packet is
transmitted over 4 contiguous slots (1 slot = 1.667 ms). The
proportional fair scheduler in each macrocell and femtocell BS
determines one user sending full-buffer traffic. The update rate of
the transmit power of the users is 150 Hz (once per 4 slots) as in
DO Rev. A. Fast fading channels (3 km/hr, 1-path) of all possible
links including both the desired and interference links are
generated using the modified Jakes model \cite{Jakes}. Effective
data rates and required per-bit signal-to-noise ratio (Eb/Nt) are
shown in Table \ref{tab:table1}. For fair comparison between the
open-loop control and the closed-loop control for $P_{max}$, the
identical interference threshold of two schemes, i.e.,
$I_{th,k^\ast}=I_{th,k^\ast}(0)$, is considered. This is achieved by
$\beta=\frac{\alpha N_0WF}{J_{k^\ast} NI_{k^\ast}(0)}$.
Here, $\alpha$ is set to be 3, and $N_0 W F$ is -109 dBm.

\subsection{Single femtocell}
Although an assumption of single femtocell per macrocell is not realistic, it provides
insight regarding the proposed algorithms' performance according to
wall penetration loss and macro-to-femto distance. To evaluate the performance of the proposed algorithms, two
performance metrics of the degradation ratio of macrocell throughput
(DRMT) $L_{T}$ and the achievement ratio of femtocell throughput
(ARFT) $A_{T}$ are defined as follows:
\begin{equation}
L_{T}=\frac{T_{m,0}-T_{m}}{T_{m,0}} ~~\text{and}~~
A_{T}=\frac{T_{f}}{T_{f,0}}, \label{eq:TputLoss}
\end{equation}
where $T_m$ and $T_{m,0}$ are the uplink average sector throughput
in the macrocell with and without the femtocell, respectively.
$T_{f,0}$ is the uplink average throughput in the femtocell when the
femtocell users employ 23 dBm of $P_{max}$, and $T_f$ is the uplink
average throughput in the femtocell using the proposed open-loop or
closed-loop power control scheme.

\begin{figure}
\begin{center}
   \includegraphics[width=4.3in]{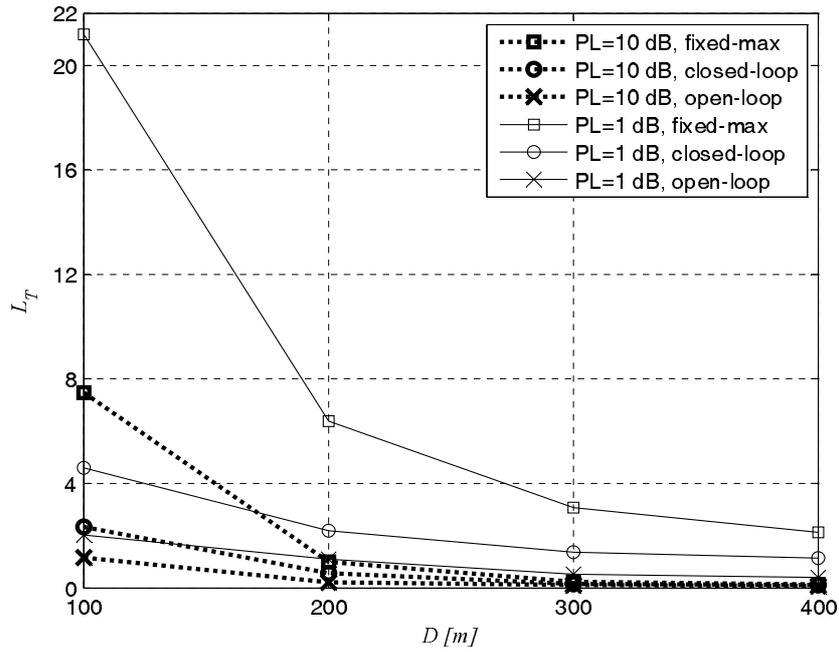}
    \caption{DRMT $L_T$ versus $D$ for the fixed maximum power transmission, the
closed-loop control, and the open-loop control schemes: $L_i$ = 0
dB}
    \label{fig:LT}
\end{center}
\end{figure}

The DRMT versus the distance between the macrocell BS and the
femtocell BS, $D$, for the fixed maximum power ($P_{max}$ = 23 dBm),
the closed-loop control, and the open-loop control schemes are
compared in Fig. \ref{fig:LT} when a external wall loss $L_e$ = 10 and
1 dB, and a internal wall loss $L_i$ = 0 dB. The DRMT decreases as
$D$ increases for all of the considered schemes as the cross-tier
interference at the macrocell BS decreases as $D$ increases. Both
the open-loop and the closed-loop control schemes are observed to
perform consistently better compared to the fixed maximum power.
Additionally, the performance gain of the two proposed schemes over
the fixed maximum power increases as $D$ and $L_e$ decrease. This is
because the cross-tier interference at the macrocell BS caused by
the fixed maximum power increases as $D$ and $L_e$ decrease, but
those with the open-loop and the closed-loop control schemes are
consistently below the threshold value, irrespective of the $D$ and
$L_e$. It is especially observed that both the open-loop and
the closed-loop control schemes achieve a DRMT of less then 0.05
regardless of the $D$ and $L_e$.

\begin{figure}
\begin{center}
   \includegraphics[width=4.3in]{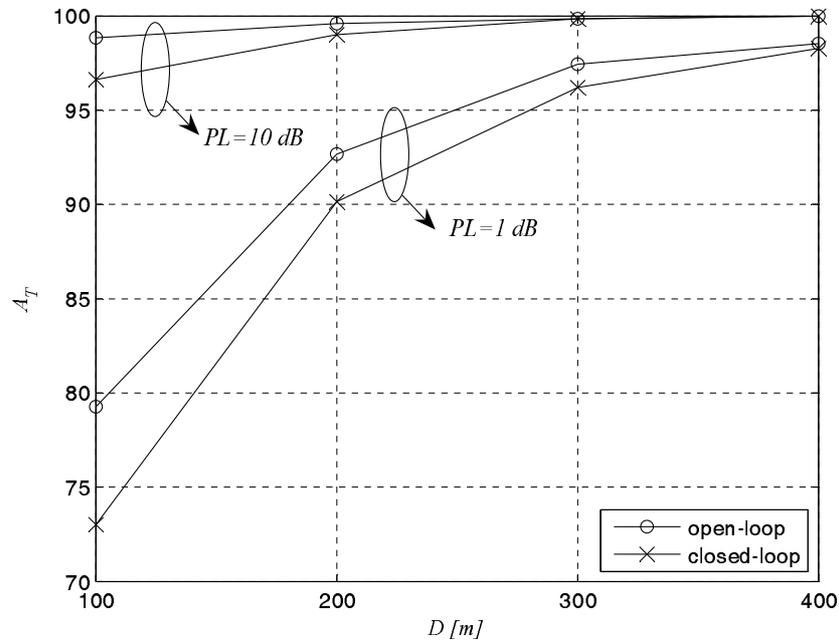}
    \caption{ARFT $A_T$ versus $D$ for the closed-loop control and the
open-loop control schemes: $L_i$ = 0 dB}
    \label{fig:AT}
\end{center}
\end{figure}

Fig. \ref{fig:AT} shows the ARFT of the open-loop and the
closed-loop control schemes as a function of $D$ when $L_e$ = 1 and
10 dB, and $L_i$ = 0 dB. The closed-loop control scheme provides
better ARFT than the open-loop control at the cost of the macrocell
throughput for all of the considered cases. However, a close
observation of Fig. \ref{fig:LT} shows that the degradation of the
macrocell throughput with the closed-loop control is minimal
compared to that of the open-loop control. The advantages of the
closed-loop control over the open-loop control increase as $D$ and
$L_e$ decrease, i.e., as the cross-tier interference dominates the
NI at the macrocell.


\subsection{Multiple femtocells}
\begin{figure}
\begin{center}
   \includegraphics[width=4.3in]{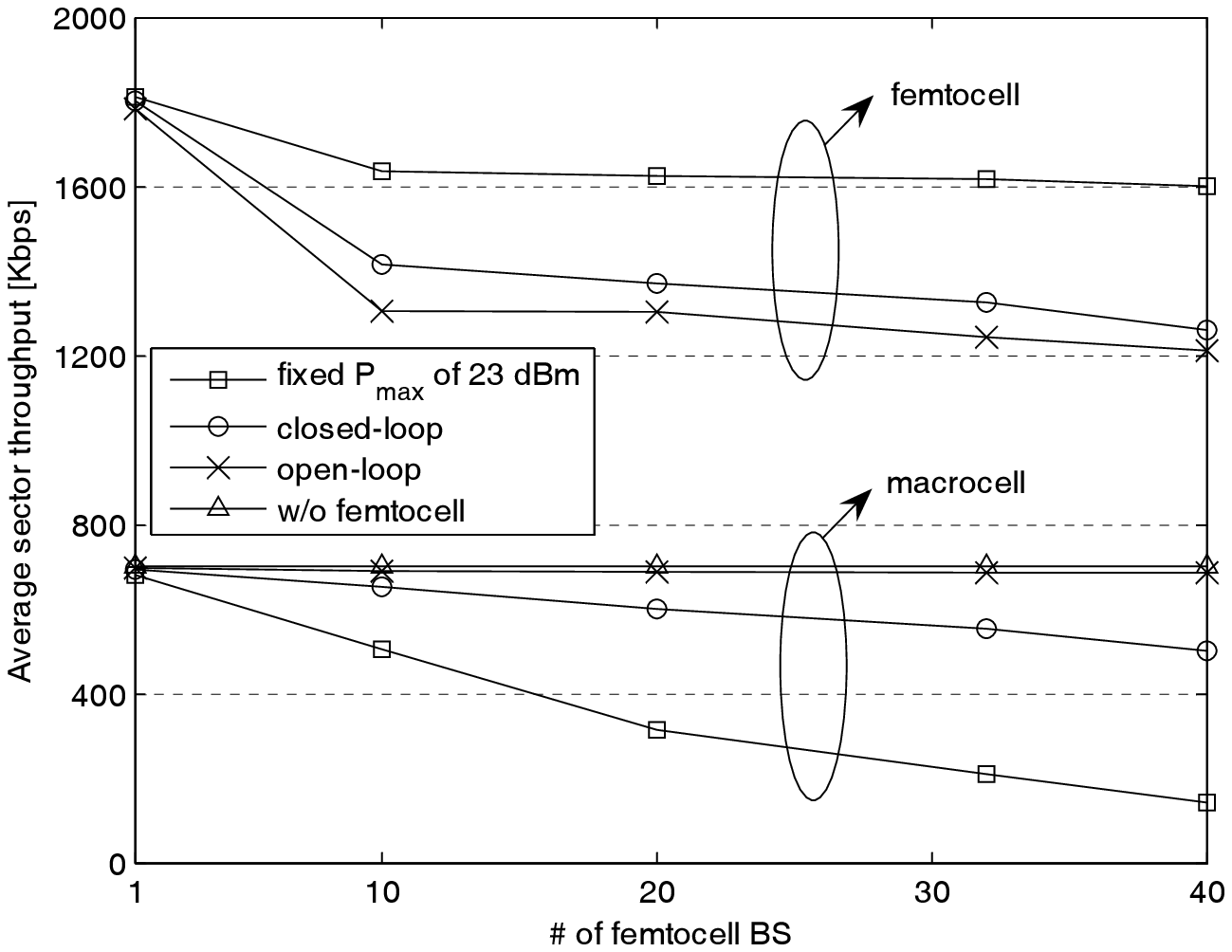}
    \caption{Average throughput versus the number of femtocell BSs per macrocell, $M$ for the closed-loop control and the
open-loop control schemes}
    \label{fig:SectorTput}
\end{center}
\end{figure}

For performance evaluation under realistic conditions, Fig.
\ref{fig:SectorTput} shows the average sector throughputs of
macrocell and femtocell as a function of the number of femtocell BSs
per macrocell, $M$. Increasing $M$ causes more severe cross-tier
interference, and thus reduces the macrocell throughput.
Furthermore, the macrocell throughput gain of the two proposed
schemes over the fixed maximum power increases as $M$ increase. This
is because the cross-tier interference at the macrocell BS caused by
the fixed maximum power scheme increases as $M$ decrease, but those
with the open-loop and the closed-loop control schemes are consistently
below the threshold value, irrespective of $M$. Fig.
\ref{fig:SectorTput} also shows that the throughput degradation of
the femtocell is smaller than that of the macrocell for a large $M$.
This is resulted from the two factors: one is that femto-to-femto
interference is smaller than the femto-to-macro cross-tier
interference due to additional wall penetration loss, and the other
is that the link loss between the femtocell user and the femtocell
BS is smaller than that between the macrocell user and the macrocell
BS.

\begin{figure}
\begin{center}
   \includegraphics[width=4.3in]{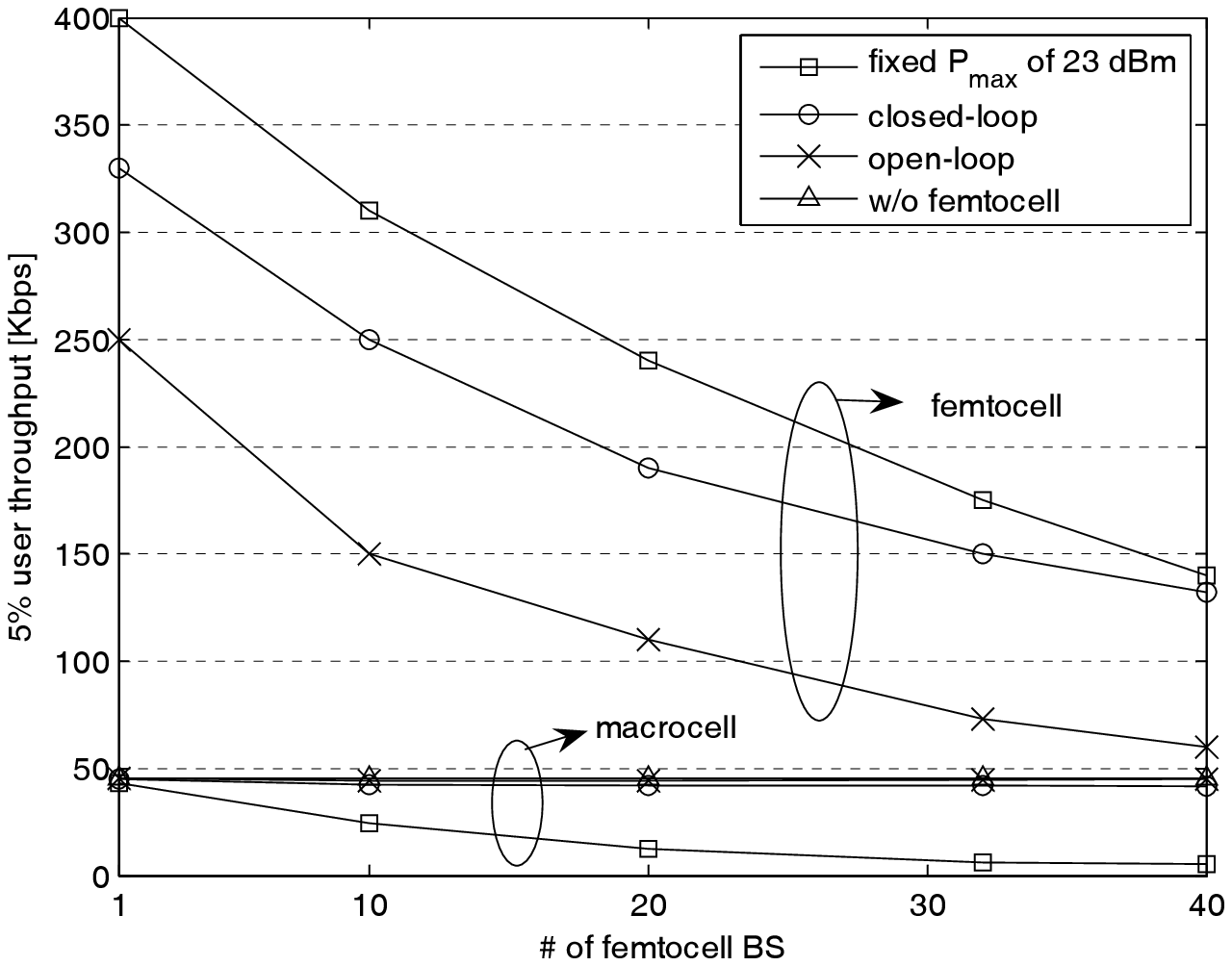}
    \caption{5\% user throughput versus the number of femtocell BSs per macrocell, $M$ for the closed-loop control and the
open-loop control schemes}
    \label{fig:EdgeTput}
\end{center}
\end{figure}

Fig. \ref{fig:EdgeTput} depicts the 5\% user throughput (defined as
the value of 5\% of the CDF of user throughput) as a function of
$M$. The 5\% user throughput of macrocell, which represents the
throughput of macrocell edge users, shows similar trend with the
average throughput of macrocell depicted in Fig.
\ref{fig:SectorTput} because the cross-tier interference at a
macrocell BS commonly affects all corresponding macrocell users.
However, the 5\% user throughput of femtocell shows different
behavior compared with the average throughput of femtocell.
Especially, the gain of the closed-loop control over the open-loop control
in terms of the 5\% user throughput of femtocell is much larger than
the gain in terms of the average throughput of femtocell.
Since the 5\% user throughput of femtocell mostly represents
the throughput of the users with low wall penetration loss and close
to the macrocell BS, in these conditions, i.e., in the cases of a
small $D$ and $L_e$, the transmit power of the femtocell user is
limited by very small $P_{max,OL}$, and thus the use of $P_{max,CL}$
leads to a higher transmit power of the femtocell user. On the other
hand, for a large $D$ and $L_e$, the transmit power is hardly
limited by $P_{max,OL}$, which is close to the maximum transmit power
of 23 dBm, and thus an increase in the transmit power is minimal
despite of the use of $P_{max,CL}$. Consequently, the femtocell
users with a small $D$ and $L_e$ are affected least by an increase
of the interference from neighbor femtocells, which result in the
higher throughput gain of the closed-loop control over the open-loop control.
Additionally, an increase in $M$ leads to a higher number of
femtocell users near the macrocell BS, which results in an increase
in the gain of the closed-loop control over the open-loop control in terms
of the 5\% user throughput of femtocell.

Fig. \ref{fig:SectorTput} and \ref{fig:EdgeTput} show that the
femtocell throughput gains of the closed-loop control over the open-loop
control are maximally 8\% and 120\% for the average throughput and
the 5\% user throughput, respectively. Considering both the
performance of macrocells and femtocells, the closed-loop control is
more effective in terms of 5\% user throughput, while the open-loop
control is good for the average throughput. On the other hand, compared with the open-loop control, the closed-loop control additionally requires for femtocell users to know the NI level at the macrocell BS, which needs additional downlink overhead for broadcasting the NI level.

\section{Conclusion}\label{sec:conclusion}
This study proposes the open-loop and the closed-loop control schemes for
the maximum transmit power of femtocell users. The two schemes
suppress the cross-tier interference under a fixed threshold and an
adaptive threshold based on the actual NI level at the macrocell BS,
respectively. Simulation results show that
both schemes effectively compensate the uplink throughput
degradation of the macrocell BS. Furthermore, the closed-loop
control scheme provides better femtocell throughput relative to the
open-loop control at a minimal cost of macrocell throughput.

\newpage

\begin{table}[tb]
\caption{Uplink data rate format} \label{tab:table1}
\begin{center}
\begin{tabular}{c|c|c}
\hline
  Payload size (bits) & Effective data rate for 4-slot (kbps) & Eb/Nt for 1\% FER (dB) \\
  \hline \hline
  128 & 19.2 & 5.6 \\ \hline
  256 & 38.4 & 5.7 \\ \hline
  512 & 76.8 & 5.8 \\ \hline
  1024 & 153.6 & 7 \\ \hline
  2048 & 307.2 & 5.4 \\ \hline
  4096 & 614.4 & 6.3 \\ \hline
  8192 & 1228.8 & 5.5 \\ \hline
  12288 & 1843.2 & 11.4 \\ \hline
\end{tabular}
\end{center}
\end{table}

%

\newpage

\newpage

\newpage

\newpage

\newpage

%
%

%

\end{document}